\documentclass[twocolumn,10pt,aps,superscriptaddress,twocolumn,showpacs,longbibliography]{revtex4}
\usepackage{epsfig,amsmath,amssymb,amsfonts,amsthm,graphicx}

\usepackage{graphicx}
\usepackage{bm}
\usepackage{hyperref}

\usepackage{ulem}
\usepackage{color}

\begin{document}

\title{Noise induced swarming of active particles}

\author{Chunming Zheng}
 \affiliation{Max Planck Institute for the Physics of Complex Systems, N{\"o}thnitzer Str. 38, 01187 Dresden, Germany}
\author{Ralf T\"onjes}%
\email{toenjes@uni-potsdam.de}
\affiliation{Institute of Physics and Astronomy, Potsdam University, 14476 Potsdam-Golm, Germany}

\begin{abstract}
We report on the effect of spatially correlated noise on the velocities of self propelled particles. Correlations in the random forces acting on self propelled particles can induce directed collective motion, i.e. swarming. Even with repulsive coupling in the velocity directions, which favors a disordered state, strong correlations in the fluctuations can align the velocities locally leading to a macroscopic, turbulent velocity field. On the other hand, while spatially correlated noise is aligning the velocities locally,
the swarming transition to globally directed motion is inhibited when the correlation length of the noise is nonzero, but smaller than the system size. We analyze the swarming transition in $d$ dimensional space in a mean field model of globally coupled velocity vectors.
\end{abstract}

\maketitle

\section{Introduction}\label{Sec:Intro}

It seems surprising that certain forms of order can arise in randomly forced systems. Yet noise and heterogeneity from different sources and at different scales is ubiquitous in nature where such synergetic effects as stochastic resonance \cite{benzi1981mechanism,Luca1998StochasticRes}, coherence resonance \cite{Haken1993CoherenceRes,Pikovsky1997CoherenceRes,Lindner2004Effects} and noise induced synchronization \cite{teramae2004robustness,goldobin2005Physica,nakao2007noise,nagai2010noise} can explain regularity despite an inherently random environment. Even in linear systems correlations in the noise can cause correlations in the system response which is famously known as Moran's theorem in ecology \cite{moran1953statistical}. In this report we study the alignment of vectors evolving on the surface of a $d$ dimensional sphere subject to polar attractive or repulsive interaction and to white noise which may be correlated between individual vectors or globally. When the vectors are interpreted as velocity vectors of self propelled active particles, for instance in the paradigmatic Vicsek model \cite{vicsek1995novel}, alignment can be observed as a macroscopic nonzero flow, resulting in an effective transport of matter and momentum. We show that a global random forcing can lead to a complete alignment of velocities, even for moderately small repulsive interaction.
In $d=2$ dimensions the vectors are characterized by a single angle, and noise induced swarming is really an expression of noise induced synchronization. Our mean field analysis in the second half of this paper generalizes this effect to higher dimensions. In the absence of coupling, and in the limit of large $d$, the restriction to the surface of the unit sphere becomes irrelevant. This is in direct analogy to the distinction between the microcanonical and the canonical ensemble in thermodynamics and the restriction to a surface of constant energy. We find that in this limit the vector components become independent linear processes and alignment of the vectors by correlated white noise is explained by Moran's theorem. In the first part of this paper, after an introduction to the Moran effect, noise induced synchronization, the similarities between the Vicsek model of active particles, the Kuramoto model and its higher dimensional generalizations, we study the effect of spatially correlated noise in a Vicsek-like dry model \cite{chate2020dry} of active particles  numerically. We observe local alignment of velocities due to the correlations in the noise, i.e. noise induced swarming. While we argue that this effect is intimately related to already known mechanisms, it has only recently been put forward as an explanation for the stabilization of small schools of fish \cite{jhawar2020noise}. The statistical physics of the swarming transition in the Vicsek model has been an active field of research for a long time (see \cite{ginelli2016physics,chate2020dry} and references therein). It is not the goal of this paper to shed light on the spatio-temporal scaling properties of the transition. Instead we demonstrate the effect of noise induced swarming by calculating velocity distributions and exact order parameters in the mean field approximation.
\\
The coefficient of correlation between two linear stochastic processes subject to additive, correlated Gaussian white noise is equal to the correlation between the two noise forces. This mathematically trivial theorem is known in ecology as the Moran effect \cite{moran1953statistical} offering an explanation for correlations in equilibrium fluctuations of species populations over large distances which are not coupled through migration. The source of correlated noise can be thought of as external perturbations, e.g. a top predator, fluctuations in a common resource or extreme weather events, acting over long distances directly or on a fast convective or diffusive time scale. The effect has also been studied in non-linear and non-equilibrium processes \cite{engen2005generalizations}, e.g. for cyclic population dynamics \cite{massie2015enhanced}. Indeed, nonlinear self-sustained oscillators may synchronize when they are subjected to correlated fluctuations. This somewhat counter intuitive behavior is known as noise induced synchronization \cite{teramae2004robustness,goldobin2005Physica,nakao2007noise,nagai2010noise} and can even be observed in chaotic oscillators subject to common noise \cite{pikovsky1992statistics}. The Kuramoto model of coupled oscillators \cite{kuramoto1975self,kuramoto1984chemical} and the Vicsek model of self propelled particles \cite{vicsek1995novel} are paradigmatic in their respective fields - synchronization and active matter. Recently, efforts have been made to exploit similarities, to generalize the respective models and to transfer results \cite{lohe2009non,tanaka2014solvable,zhu2013synchronization,crnkic2021synchronization,markdahl2020high,Levis2019activity,maistrenko2018spp,lipton2021kuramoto,chepizhko2010relation,degond2014hydrodynamics,chandra2019continuous,zheng2021transition}. Oscillations in the Kuramoto model are naturally related to vortices in active matter flows and to helical motion \cite{chandra2019continuous,zheng2021transition,pikovsky2021transition}. The Vicsek model, on the other hand, is easily formulated in three dimensional space and higher dimensional generalizations of the Kuramoto model have been proposed recently. 

Statistical thermodynamics of active particles, elucidating the origin and the often dominating role of fluctuations in microscopic non-equilibrium systems, is another active field of research \cite{bechinger2016active,bain2019dynamic,fodor2021active,fodor2016AOUP}. However, in the Langevin description of microscopic stochastic dynamics the fluctuations are often assumed to be independent Gaussian white noise. Allowing the noise to be autocorrelated in time can give rise to novel effects \cite{fodor2016AOUP,toenjes2010colored,nagai2015Memory}. Here we study the effect of spatial correlations which has not been considered so far. In two dimensions the analysis of noise induced synchronization can directly be applied to the Vicsek model where it manifests as noise induced swarming, as we refer to the emergent alignment of velocity vectors under the influence of common or correlated noise. Indeed, common multiplicative noise in the form of finite size fluctuations has recently been identified as a mechanism to stabilize coherent swarming in small schools of fish \cite{jhawar2020noise}.

In the next section Sec.\ref{Sec:Spatial} we will introduce Langevin equations for a Vicsek-like dry model of active particles \cite{chate2020dry}, i.e. polar particles without hydrodynamic equations for the medium through which the particles interact. We study noise induced swarming in this model with polar interaction in two and three dimensions numerically and find that spatial correlations in the noise lead to an increase in the local alignment of velocities at the scale of the correlation length but inhibits global synchronization. In Sec.\ref{Sec:MeanField} we analyze the model in the mean field approximation and find the distributions of order parameters for arbitrary dimensions $d$. The alignment of high dimensional vectors confined to the surface of a hypersphere is of interest in opinion dynamics and consensus based optimization \cite{caponigro2015opinions,fornasier2021consensus}. It is in the limit of high dimensions that the formal connection between noise induced swarming and Moran's theorem becomes apparent. 
\section{Spatially extended model}\label{Sec:Spatial}
The original Vicsek model of self propelled particles \cite{vicsek1995novel} defines a time-discrete map for the positions $\vec{x}_n\in\mathbb{R}^d$ and unit length velocities $\vec{v}_n\in S^{d-1}$, $v_n=|\vec{v}_n|=1$ of particles $n=1\ldots N$ in a $d=2$ two dimensional system. In each time step the positions change according to the velocities and the velocities assume  the direction of the average velocity within a coupling range $R$ plus a uniformly distributed individual random angle. Here we adopt a more mechanical model \cite{chate2004onset}, where so called vectorial noise acts in the same way as the coupling forces. The dynamics of the self propelled particles is given by Langevin stochastic differential equations, i.e. continuous in time and subject to Gaussian white noise. Forces,  including the random fluctuations, act only in the orthogonal directions on the velocities, thus changing the direction but not the speed of the particles. In units of time and space where 
$v=1$
the equations of motion are
\begin{eqnarray}
        \dot{\vec{x}}_n &=& \vec{v}_n \label{Eq:VicsekModelX}\\
        \dot{\vec{v}}_n &=& \vec{F}_n -\left(\vec{F}_n\cdot\vec{v}_n\right)\vec{v}_n \label{Eq:VicsekModelV}
\end{eqnarray}
i.e. $\dot{\vec{v}}\cdot\vec{v}=0$. Any common component in the forces $\vec{F}_n$, including a common noise source, will contract all velocities on the surface of the sphere into that direction, may overcome independent noise and heterogeneity in the forces and lead to an emergent collective swarming state. The assumption of a constant velocity, identical for all particles, is a strong simplification, implying low heterogeneity of the particles and a fast relaxation to the terminal velocity where friction and propulsion forces are balanced. The force acting on a particle is aligning it to the local velocity field with a coupling strength $K$ and has a stochastic component which we model as a Gaussian white noise field $\vec{\eta}(\vec{x},t)$ of strength $D$
\begin{equation}\label{Eq:LocalForce}
    \vec{F}_n =  K \left\langle \vec{v}_n\right\rangle_R + \sqrt{2D}\vec{\eta}(\vec{x}_n,t).
\end{equation}
The term $\left\langle \vec{v}_n\right\rangle_R$ denotes the average velocity of particles within the coupling distance $R$ to the  position $\vec{x}_n$. 
\\ \\
The projection  in \eqref{Eq:VicsekModelV} makes the noise multiplicative. 
The stochastic differential equation \eqref{Eq:VicsekModelV} has therefore to be interpreted in the sense of Stratonovich to ensure the constant velocity amplitude. Unlike the contraction into the direction $\vec{F}(t)$ in \eqref{Eq:VicsekModelV}, a random but common rotation, so called angular noise, would not change the angles between velocities. 
Only correlations of the vectorial noise in $\vec{F}$ will lead to noise induced swarming. Let us define the set of neighbors $U_n(R) = \lbrace m : |\vec{x}_m-\vec{x}_n|\le R\rbrace$ including particle $n$, the number of neighbors $k_n(R)=|U_n(R)|$, the local velocity field
\begin{equation}\label{Eq:LocalVelo}
    \vec{V}_n = \left\langle \vec{v}_n\right\rangle_R = \frac{1}{k_n(R)}\sum_{m\in U_n(R)} \vec{v}_m
\end{equation}
as well as the global order parameter $V = \frac{1}{N}\left|\sum_n \vec{v}_n\right|$ and the average local order parameter $V_R=\frac{1}{N} \sum_{n}|\vec{V}_n|$.
Spatially correlated but not necessarily identical noise introduces another time and length scale into the model, i.e. the coefficient of correlation $0\le q\le 1$ (we do not consider anti-correlated noise) and a correlation length $\Lambda$. In our numerical simulation we create spatially correlated noise by averaging distributed independent white noise sources. Indeed, assigning an independent white noise source $\vec{\xi}_n=(\xi_{ni})$ with $n=1\ldots N$, $i=1\ldots d$ and $\langle \xi_{mi}(t)\xi_{nj}(t')\rangle = \delta_{mn}\delta_{ij}\delta(t-t')$ to each particle, and defining
\begin{equation} \label{Eq:LocalNoise}
    \vec{\eta}(\vec{x}_n,t) = \sqrt{k_n(\Lambda)}\left\langle \vec{\xi}_n\right\rangle_\Lambda = \frac{1}{\sqrt{k_n(\Lambda)}}\sum_{m\in U_n(\Lambda)} \vec{\xi}_m,
\end{equation}
we obtain spatially correlated white noise $\vec{\eta}(\vec{x}_n,t) = \vec{\eta}_n = (\eta_{ni})$ with
\begin{equation}\label{Eq:NoiseCorrel}
    \langle \eta_{mi}(t)\eta_{nj}(t')\rangle = q(\vec{x}_m,\vec{x}_n) \delta_{ij}\delta(t-t')
\end{equation}
and with a spatial correlation function of characteristic length scale $\Lambda$
\begin{equation}\label{Eq:DenseCorrel}
    q(\vec{x}_m,\vec{x}_n) = \frac{|U_m(\Lambda)\cap U_n(\Lambda)|}{\sqrt{k_m(\Lambda)k_n(\Lambda)}}.
\end{equation}
The specific source of the correlations and shape of the correlation function is not essential for the effect of noise induced swarming but may effect the statistics of turbulent states. Power-law correlations in the noise would lead to power-law correlations in the velocity field and vanishing difference in the noise at small spatial distances may lead to the formation of separated clusters in the long time limit. Note  that uncorrelated noise will, in the thermodynamic limit of large $N$, affect macroscopic observables smoothly on diffusive time and length scales. On the other hand, spatial correlations in the noise are macroscopic forces, leading to macroscopic fluctuations of observables at the scale of the correlation length, even in the thermodynamic limit.
%
\begin{figure}[t]
\setlength{\unitlength}{1cm}
\begin{picture}(4.2,4.2)
\put(-0.05,0){\includegraphics[height=4.3cm]{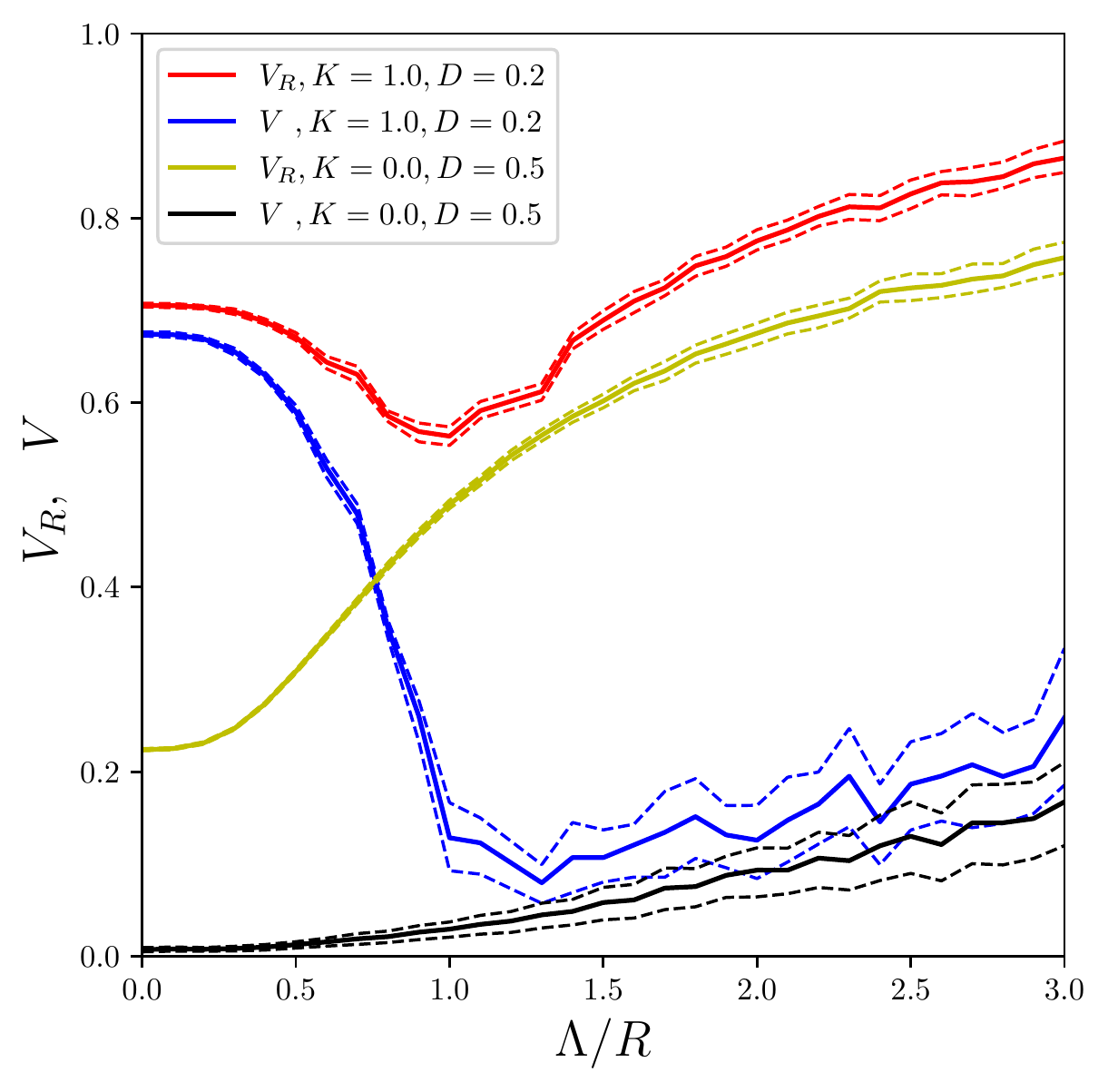}}
\put(-0.2,3.8){\bf (a)}
\end{picture}
\begin{picture}(4.2,4.2)
\put(0,0.3){\includegraphics[height=4.0cm]{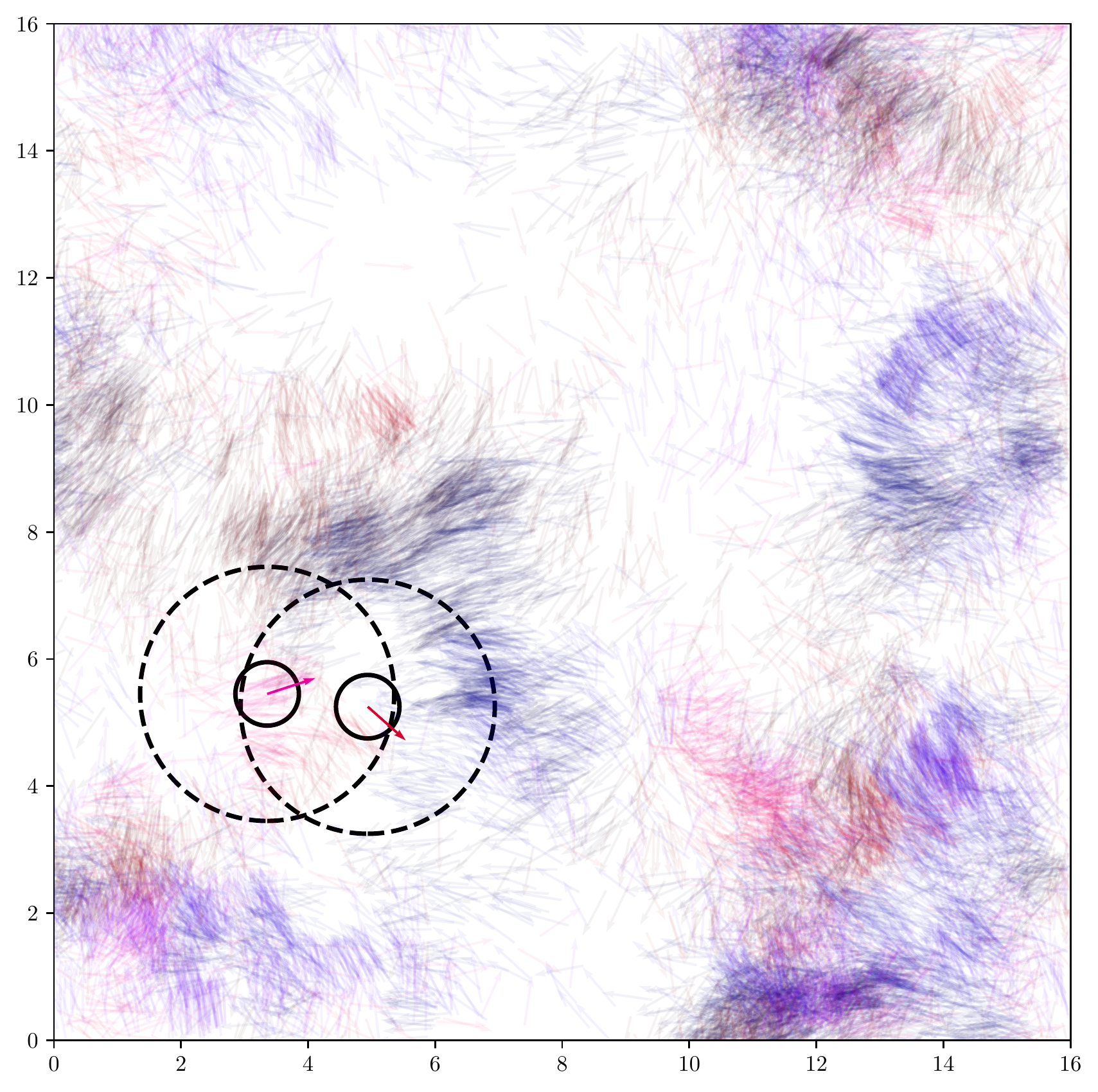}}
\put(0.2,3.8){\bf (b)}
\end{picture}
\begin{picture}(4.2,4.6)
\put(0,0.3){\includegraphics[height=4.4cm]{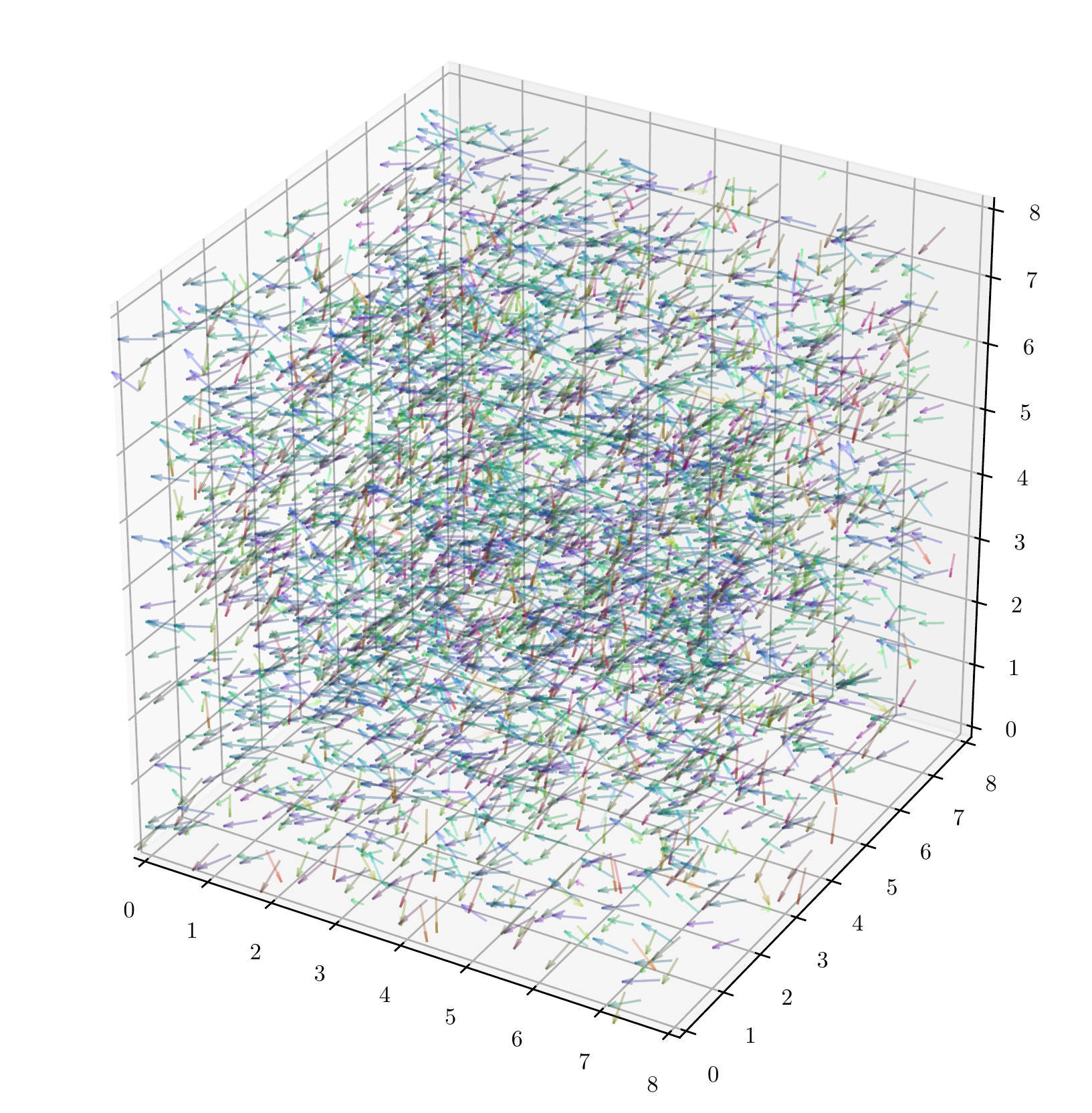}}
\put(0.2,3.8){\bf (c)}
\end{picture}
\begin{picture}(4.2,4.6)
\put(0,0.3){\includegraphics[height=4.4cm]{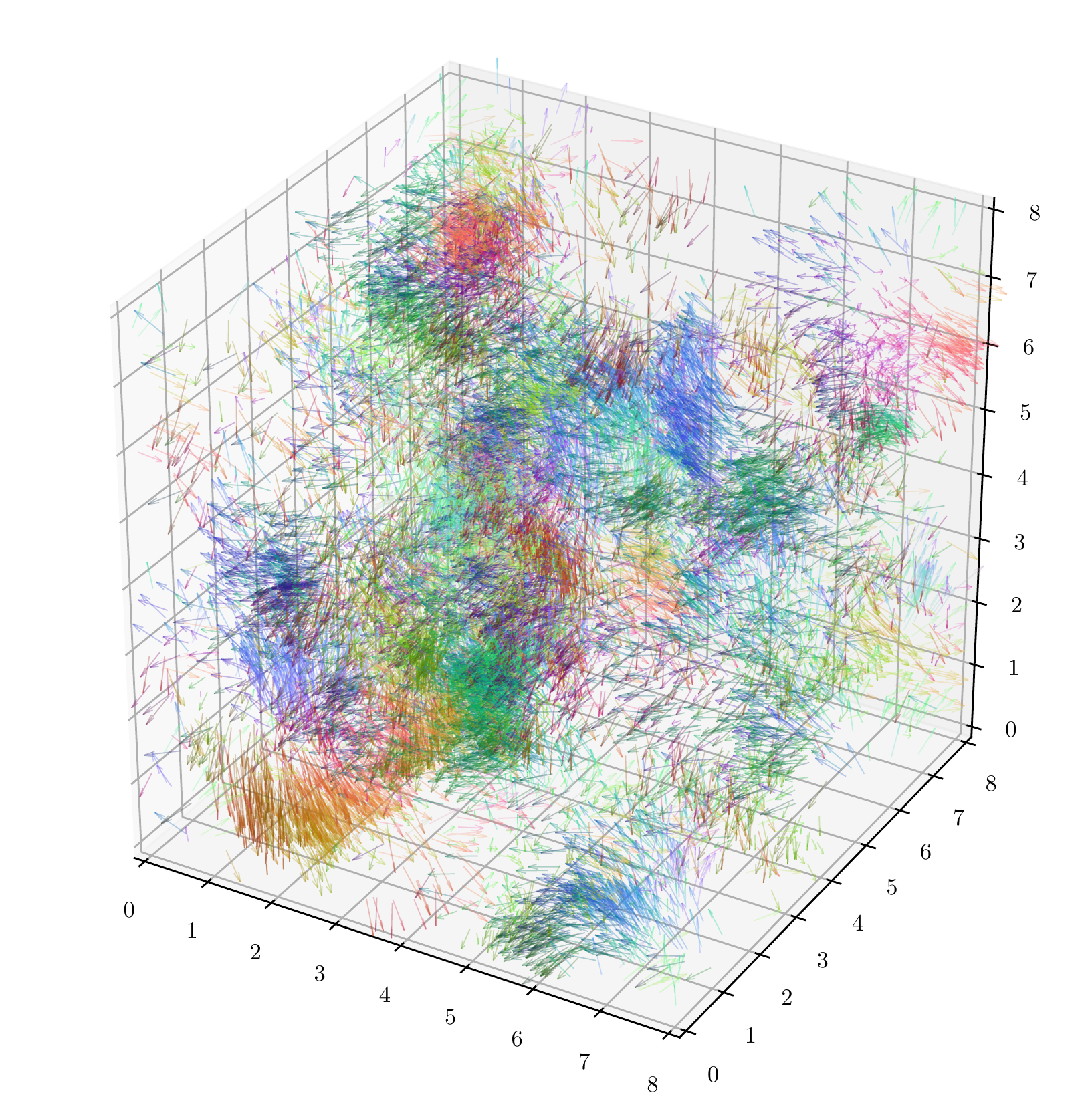}}
\put(0.2,3.8){\bf (d)}
\end{picture}
\caption{The local order $V_R$ is increasing when the noise correlation length $\Lambda$ is increased, while global alignment $V$ is inhibited. (a) Local and global order parameters $V_R$ and $V$ in the 3d model of $N=16384$ self propelled particles in a periodic domain of side length $L=8$, with coupling distance $R=0.5$ and different noise correlation lengths $\Lambda$. Solid lines are median values and dashed lines $\pm 25\%$ percentiles. Red and blue lines for coupling strength $K=1.0$ and noise strength $D=0.2$. Yellow and black lines for uncoupled particles with $K=0.0$ and $D=0.5$. (b) Turbulent noise induced swarming state. Velocity vectors of $N=16384$ particles in a 2d periodic domain of side length $L=16$, zero coupling strength $K=0$ over a coupling radius $R=0.5$ (solid black circles) and noise strength $D=0.5$ with noise averaging over circles of radius $\Lambda=2$ (dashed circles). The shades of the vectors indicate the direction. In (c) and (d) we show the velocity vectors of $N=16384$ particles (only 4000 are shown in (c)) in a periodic 3d medium with $L=8$. Coupling radius is $R=0.5$, coupling strength and noise strength are $K=1.0$ and $D=0.2$ with correlation lengths (c) $\Lambda=10^{-4}$ and (d) $\Lambda=1.0$. Globally directed swarming state with $V\approx 0.66$ in (c) and turbulent velocity field due to correlated noise in (d). Velocity directions are also indicated by color}\label{Fig:Fig01}
\end{figure}
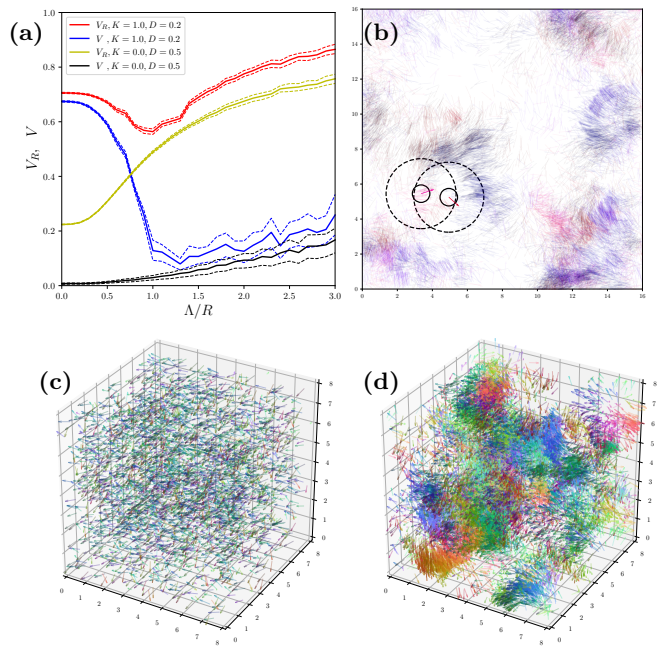
In Fig.\ref{Fig:Fig01} we demonstrate the emergence of a macroscopic irregular velocity field, characterized by high local order and low global order, through noise induced swarming. We call a velocity field turbulent in this sense. We have simulated $N=16384$ particles in two and three dimensional domains with periodic boundary conditions and side lengths $L=16$ and $L=8$, respectively. The particle velocity is $v=1$, and the coupling radius is $R=0.5~$. We use the Euler-Maruyama method with additional renormalization of the velocity vectors after each step ($dt=0.01$) to integrate the Langevin equations \eqref{Eq:VicsekModelX}-\eqref{Eq:NoiseCorrel}. 
In panel (a) the local and global order parameters in $d=3$ dimensions are shown as a function of the correlation length scale $\Lambda/R$. Without coupling, i.e. $K=0$ (Fig.1a black and yellow curves) and for spatially uncorrelated noise $\Lambda\to 0$, the velocities are independent and uniformly distributed on the unit sphere resulting in a low global order parameter $V=1/\sqrt{N}\approx0.008$ and a moderate local order parameter $V_R=1/\sqrt{k(R)}\approx 0.24$. Increasing the correlation length the particle velocities at distances $O(\Lambda)$ become correlated but are uncorrelated over longer distances. The local order parameter $V_R$ increases, while the global order parameter $V$ remains low. When $K$ is large enough to force velocity alignment in the Vicsek model with uncorrelated noise of strength $D$ (Fig.\ref{Fig:Fig01}a blue and red curves and Fig.\ref{Fig:Fig01}c), increasing the correlation length but keeping the noise strength constant, can destroy the state of globally directed motion (Fig.\ref{Fig:Fig01}a,d). Panel (b) shows the turbulent $d=2$ velocity field for uncoupled particles ($K=0$) subject to white noise of strength $D=0.5$ and noise correlation length scale $\Lambda=2$. Panels (b) and (d) show states of active matter turbulence, in Fig.\ref{Fig:Fig01}b without coupling (pure noise induced swarming) and in Fig.\ref{Fig:Fig01}d with attractive coupling. Repulsive coupling will only further decrease the average local order parameter.
\section{Mean field analysis}\label{Sec:MeanField}
If $R$ is larger than the system size and the correlation function is approximately constant, the global order parameter is described by mean field theory \cite{escaff2020flocking,chepizhko2009kinetic}. For finite coupling radius, if the timescale of the local alignment is faster than the timescale of the motion, only
the local velocity fields may be described by mean field theory. 
It is noteworthy that there is no explicit density dependent interaction in this model, such as volume excluding repulsive forces or cohesive interaction. Fluctuations in the density larger than the expected finite size fluctuations are an emergent effect. Large flocks of particles going in the same direction at the same speed stay together longer and grow through assimilation, while they also may break through scattering. Such coherent structures play an important role in the propagation and eventual divergence of velocity correlations through coupling \cite{chate2004onset}.
Since mean field analysis assumes high particle densities we simulate rather small spatial domains and moderately high densities ($N/L^3=32$ in 3d and $N/L^2 \approx 64$ in 2d). However, mean field analysis can only locally predict noise induced swarming and makes no prediction for the transition to global alignment in spatially extended systems. If the particle velocities are subject to a global mean field force and noise
\begin{equation}\label{Eq:MF_Force}
    \vec{F}_n = K\langle \vec{v} \rangle + \sqrt{2D}\vec{\eta}_n(t)
\end{equation}
with correlated Gaussian white noise $\langle \eta_{mi}(t)\eta_{nj}(t')\rangle=[q(1-\delta_{nm})+\delta_{nm}]\delta_{ij}\delta(t-t')$,
then the order parameter $V$ of the system depends only on the relative coupling strength $\kappa = K/D$ and the coefficient of correlation $0\le q\le 1$ (Fig. \ref{Fig:Fig02}a). In the following we present exact expressions for the order parameter $V$ with purely uncorrelated noise ($q=0$) and with identical noise ($q=1$) and for the velocity correlation $C=\vec{v}\cdot\vec{v}'$ in uncoupled oscillators ($\kappa=0$) but arbitrary noise correlation $q$. 
\\ \\
The first case of uncorrelated noise results in a Boltzmann type stationary distribution \cite{kirkpatrick2016asymptotics,zheng2021transition}
\begin{equation}\label{Eq:Boltzmann}
    p(\vec{v}) = \frac{1}{Z} e^{\kappa \vec{V}\cdot\vec{v}}
\end{equation}
with a normalization constant $Z$, where the order parameter is implicitly given \cite{kirkpatrick2016asymptotics,SM} as a ratio of modified Bessel functions of the first kind
\begin{equation}\label{Eq:ImplicitV}
   V = \frac{I_{d/2}(\kappa V)}{I_{d/2-1}(\kappa V)}.
\end{equation}
The bifurcation curve $V=V(\kappa)$, shown in Fig.\ref{Fig:Fig02}b and compared to simulations of the Langevin Equations, has the parametric form $V(x)=I_{d/2}(x)/I_{d/2-1}(x)$, $\kappa(x)=x/V(x)$ and $x \ge 0$. At the critical coupling $\kappa_{cr}=d$, where $x\to 0$, the order parameter becomes zero with square root scaling \cite{zheng2021transition}. Interestingly a similar second order transition is observed in active Ornstein-Uhlenbeck processes with non-linear directional coupling \cite{dossetti2015emergence}.
\\ \\
With only common noise, i.e. $q=1$, all velocities are subject to the same force $\vec{F}_n(t)=\vec{F}(t)$. In analogy to the invariant Ott-Antonsen manifold for phase oscillators forced in the first harmonics \cite{ott2008low} a family of continuous distributions on higher dimensional unit spheres given by the hyperbolic Poisson kernel 
\cite{chandra2019complexity}
\begin{equation}\label{Eq:PoissonKernel}
    p(\vec{v}) = \frac{1}{Z}\left(\frac{1-a^2}{|\vec{v}-\vec{a}|^2}\right)^{d-1}
\end{equation}
exists, which includes a uniform initial distribution ($a=0$) and, under common forcing, is invariant under the flow defined by \eqref{Eq:VicsekModelV}. The ensemble mean velocity is a function of the parameter $\vec{a}$ 
with $|\vec{a}|=a\le 1$
\begin{equation}\label{Eq:v_nonlinear}
    \langle\vec{v}\rangle = \Phi_d(a)\vec{a},
\end{equation}
$0<\Phi_d(a)\le 1$ and $\Phi_d(1)=1$.
Only for $d=2$ we have $\Phi_2(a)=1$ and the mean velocity is equal to $\vec{a}$. The functions $\Phi_d(a)$ are related through recurrences and have expressions which increase in complexity with the dimension $d$ \cite{Kato2020Cauchy,crnkic2021synchronization}. However, the parameter $\vec{a}$ follows a simple dynamics inside the $d$-dimensional unit sphere \cite{chandra2019complexity}
\begin{equation}\label{Eq:LowDimDyn_a}
    \dot{\vec{a}} = \frac{1}{2}(1+a^2)\vec{F} - \left(\vec{F}\cdot\vec{a}\right)\vec{a}.
\end{equation}
After calculating the drift and diffusion coefficients in the Fokker Planck equation of $a=|\vec{a}|$ for this Stratonovich Langevin equation, we can formally write down the stationary distribution (Appendix A)
\begin{equation}\label{Eq:C2_dist}
    p(a) = \frac{1}{Z}(1-a^2)^{-d}a^{1-d}\exp\left[\int^a \frac{2\kappa\Phi_d(s)s}{(1-s^2)}ds\right].
\end{equation}
The time average of the ensemble mean velocity $\langle V\rangle_t = \left\langle\Phi_{d}(a)a\right\rangle$ can be calculated either numerically or, for $d=2$ and $d=4$, as a ratio of special functions (Appendix A).
For $d=2$ 
\begin{equation}\label{Eq:V_common_2d}
    \langle V\rangle_t = \frac{1}{2}B\left(\frac{1}{2},-\kappa\right)
\end{equation}
with the Beta function $B(a,b)$ and for $d=4$ 
\begin{equation}\label{Eq:V_common_4d}
    \langle V\rangle_t = -\frac{3}{2\kappa}\,\frac{Z_4(3,3;\kappa)+2Z_4(5,4;\kappa)}{Z_4(4,4;\kappa)}
\end{equation}
with
\begin{equation}\label{Eq:V4d_Z4}
    Z_4(a,b;\kappa) = M\left(\frac{a}{2},1-b+\frac{a}{2}-\kappa;\frac{\kappa}{2}\right)B\left(\frac{a}{2},1-b-\kappa\right)
\end{equation}
and Kummer's confluent hypergeometric function $M(a,b;z)$. We plot the analytic expressions \eqref{Eq:V_common_2d}-\eqref{Eq:V4d_Z4} for $d=2$ and $d=4$, as well as the numerical evaluation of the mean velocity from \eqref{Eq:C2_dist} for $d=3$, together with the data obtained in the simulation of the Langevin equations in Fig.\ref{Fig:Fig02}c.
We observe, that with $\lim_{a\to 1}\Phi_{d}(a)=1$ 
the density \eqref{Eq:C2_dist} is normalizable at the pole $a=1$ only 
if $\kappa<1-d$, i.e. $\kappa_{cr}=1-d$.
For larger values the common noise leads to a complete alignment even for negative coupling $1-d<\kappa<0$. For repulsive coupling in the range $-d<\kappa<1-d$ the probability density is normalizable but divergent at $a=1$. This regime 
is characterized by intermittent strong synchronization and desynchronization.
\\ \\
The case of purely noise induced swarming with $K=0$ can be analyzed in a similar way. The square of the mean velocity is given by the ensemble average of the velocity correlation $C=\vec{v}\cdot\vec{v}\,'$
\begin{equation}\label{Eq:V2_general}
    V^2 = \frac{1}{N^2}\sum_{n,m} \vec{v}_m\cdot\vec{v}_n.
\end{equation}
With correlated noise this ensemble average is fluctuating in time but the time average is given by the expected value of $C$ with respect to the stationary distribution (Appendix A)
\begin{equation}\label{Eq:CorrelDist}
    p(C) = \frac{1}{Z}\frac{(1-C^2)^{\frac{d-3}{2}}}{(1-qC)^{d-1}}.
\end{equation}
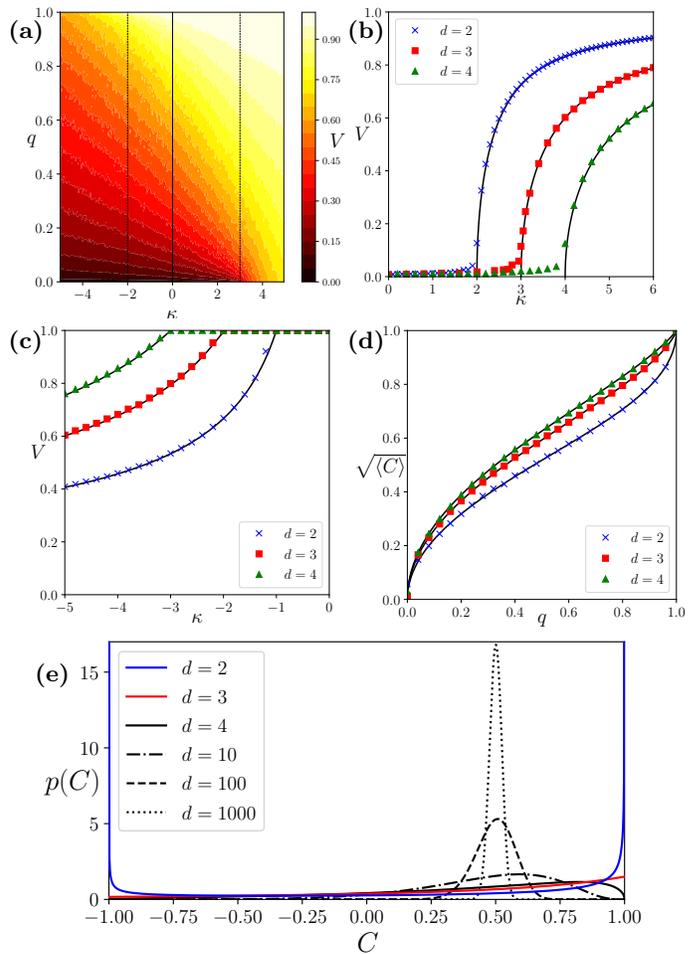
\begin{figure}[t]
\setlength{\unitlength}{1cm}
\begin{picture}(4.2,4.2)
\put(-0.05,-0.2){\includegraphics[height=4.4cm]{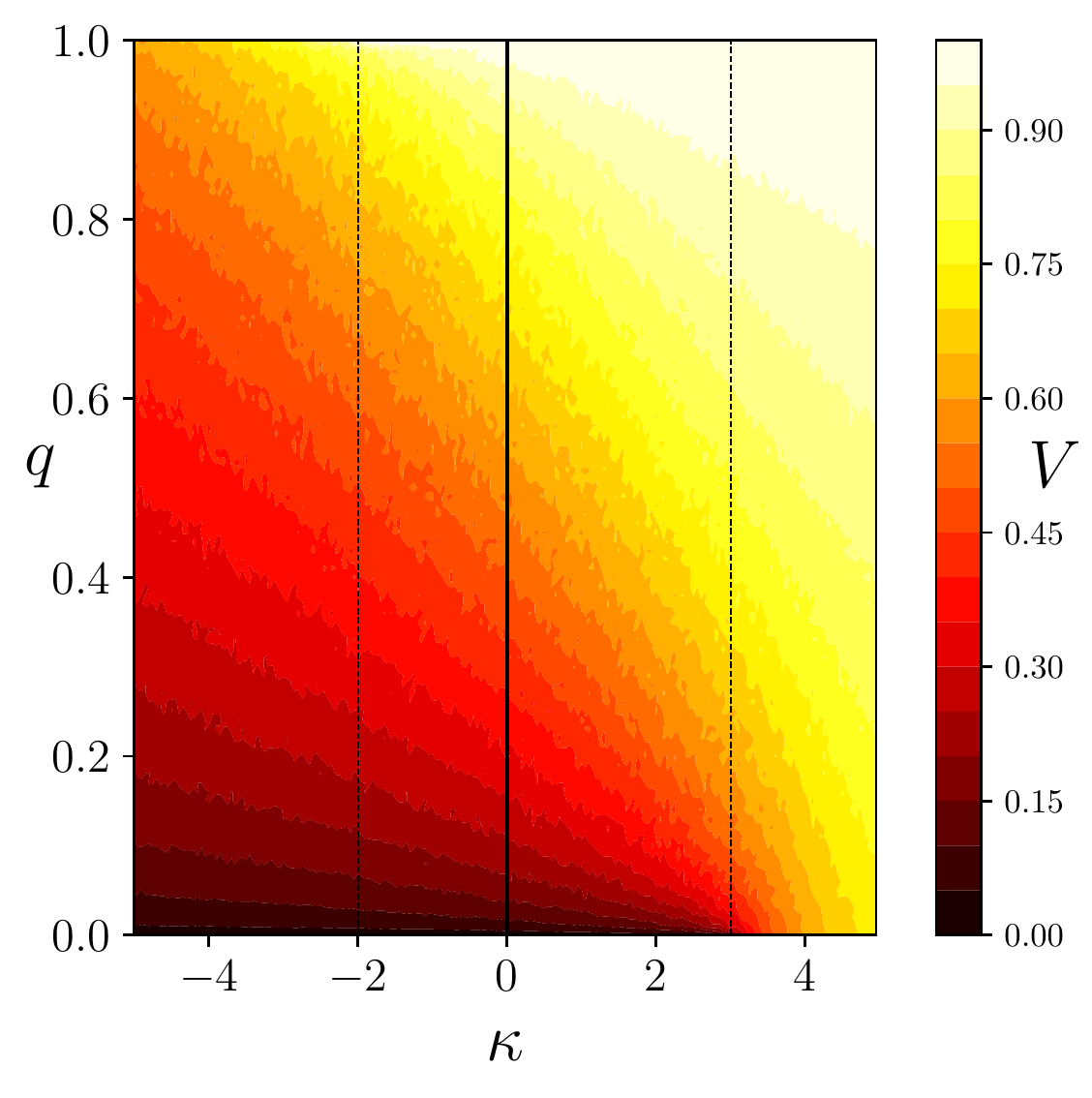}}
\put(-0.2,3.7){\bf (a)}
\end{picture}
\begin{picture}(4.2,4.2)
\put(0,0){\includegraphics[height=4.2cm]{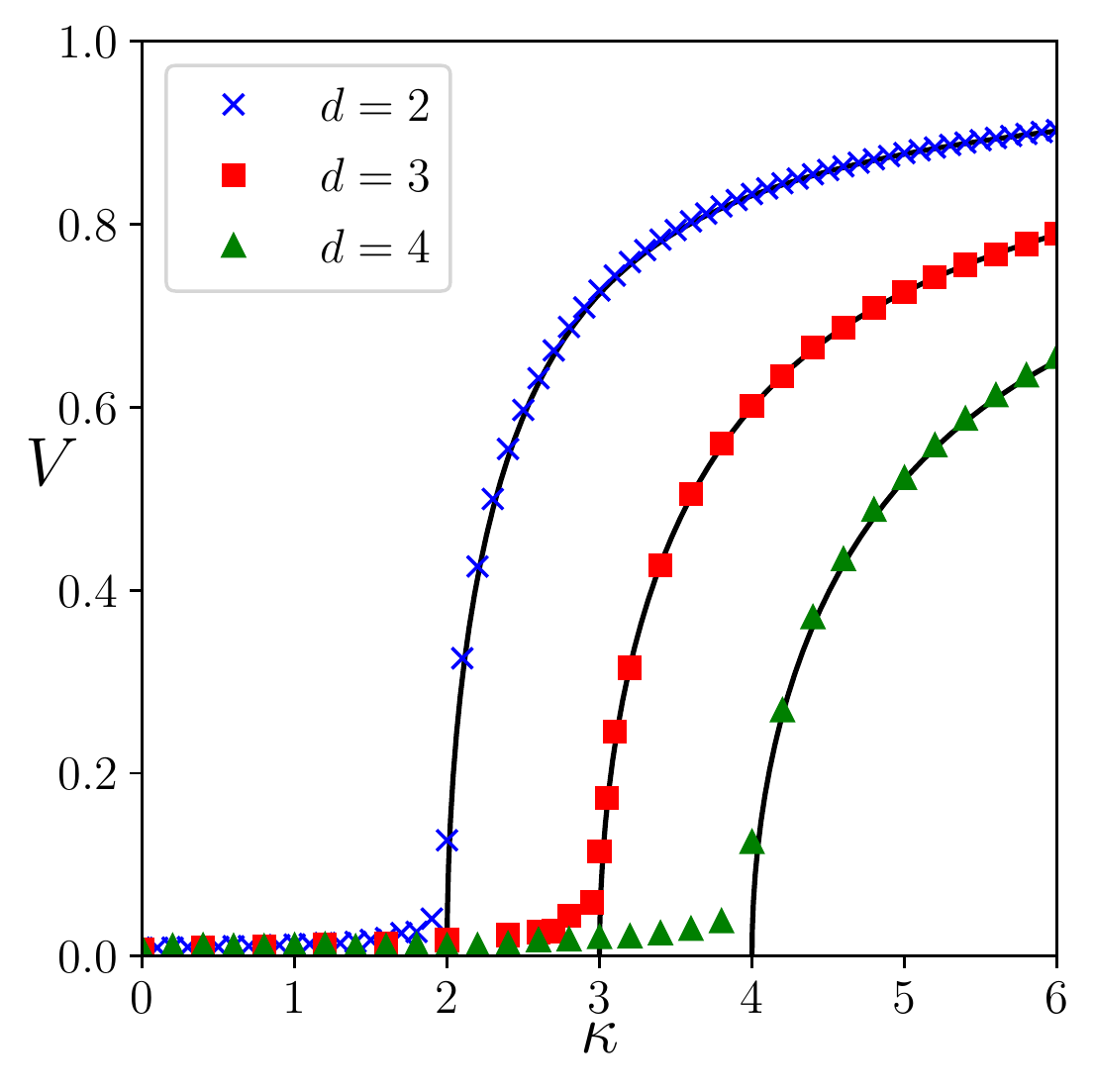}}
\put(0,3.7){\bf (b)}
\end{picture}
\begin{picture}(4.2,4.2)
\put(0,0){\includegraphics[height=4.2cm]{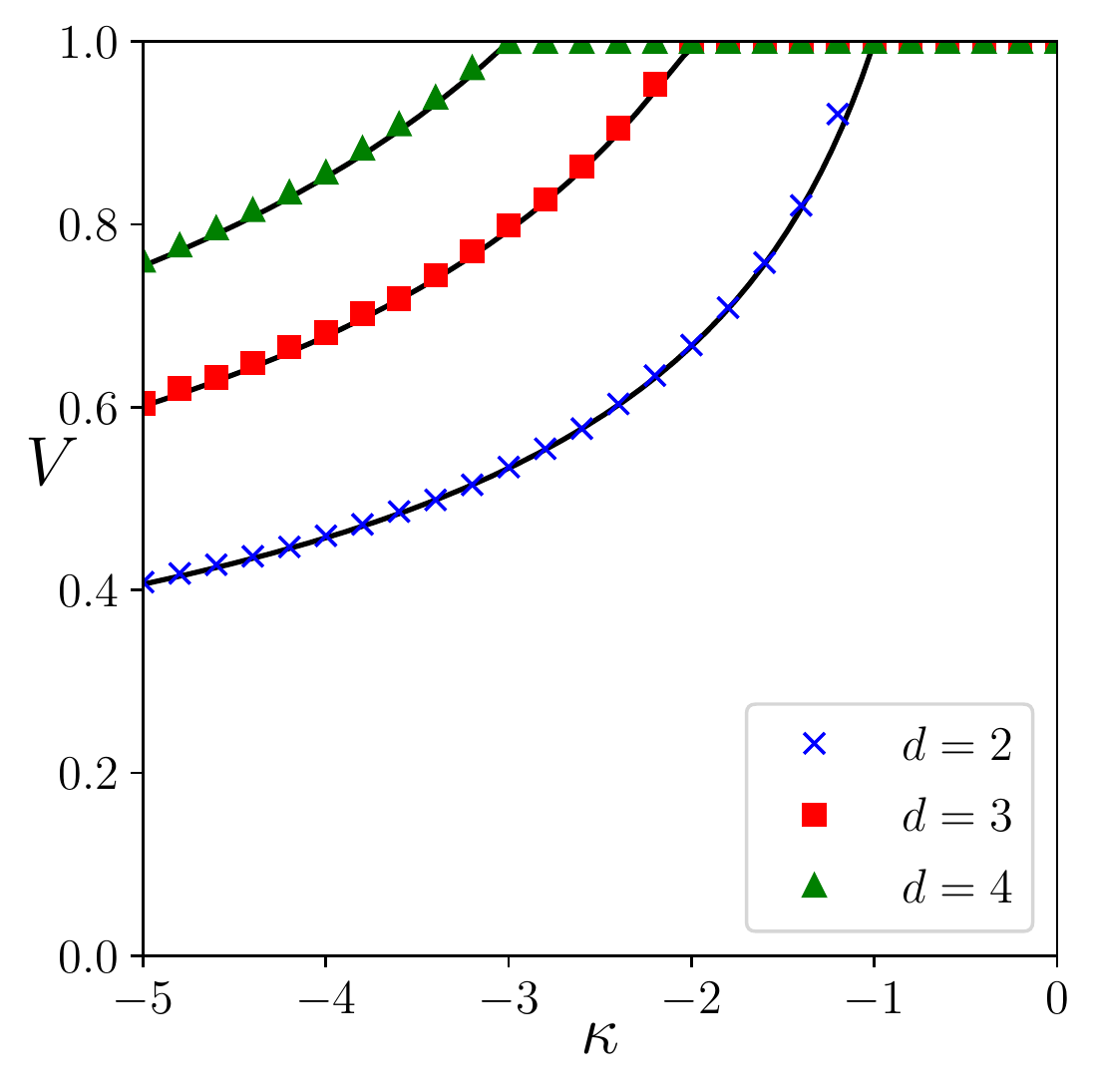}}
\put(-0.2,3.8){\bf (c)}
\end{picture}
\begin{picture}(4.2,4.2)
\put(0,0){\includegraphics[height=4.2cm]{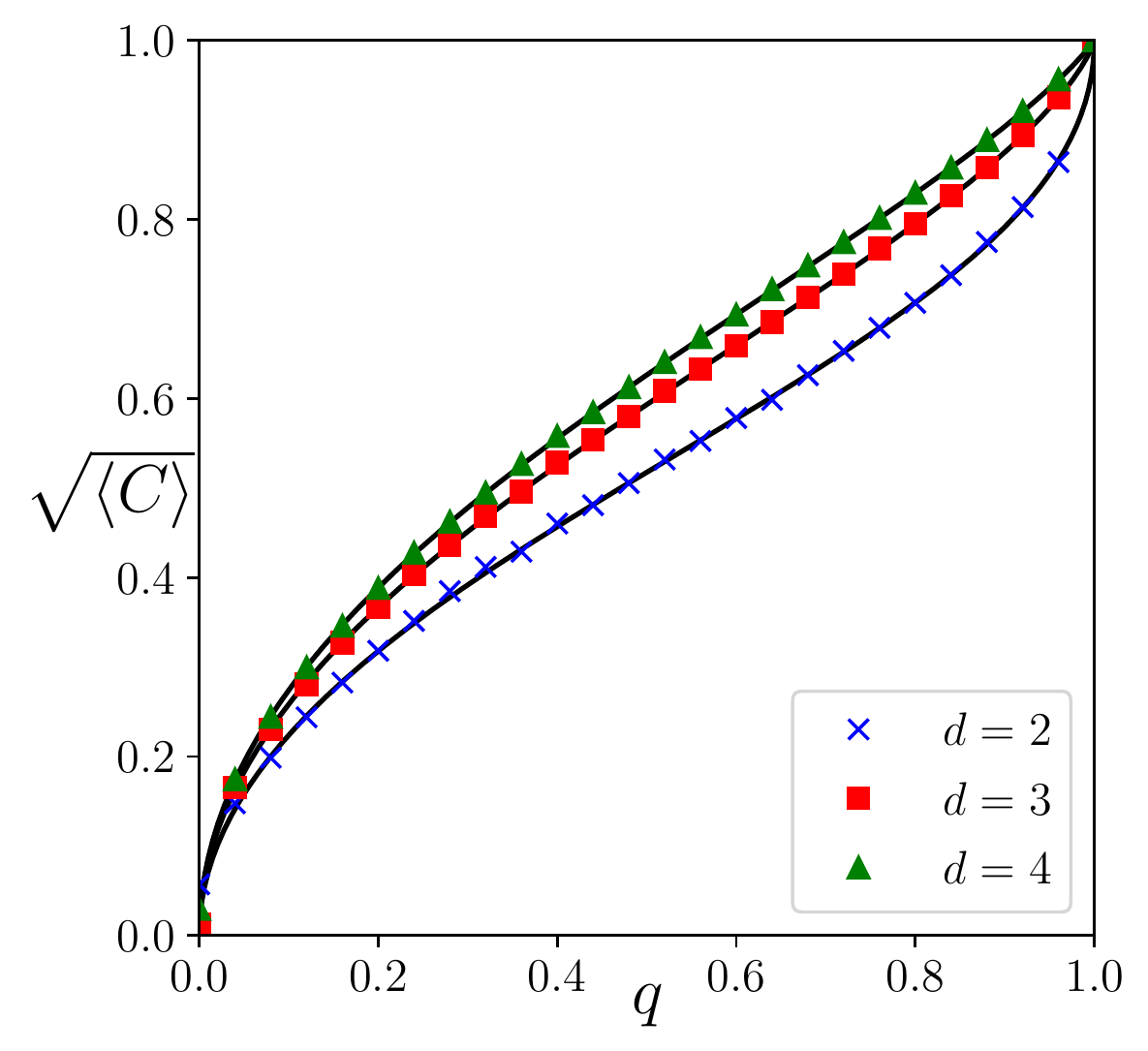}}
\put(0,3.8){\bf (d)}
\end{picture}
\begin{picture}(8.2,4.4)
\put(0,0){\includegraphics[width=8.2cm]{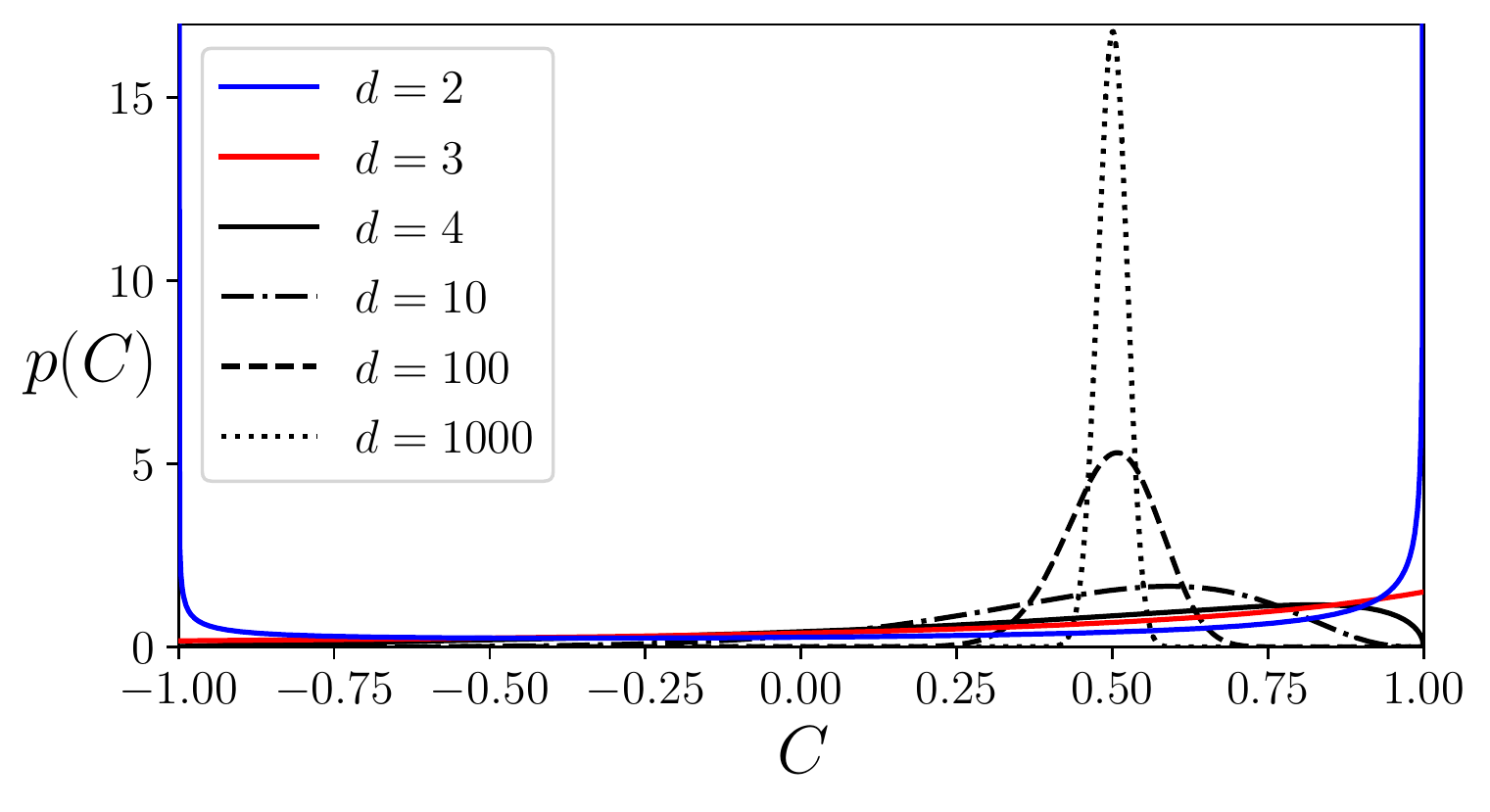}}
\put(0,3.8){\bf (e)}
\end{picture}
\caption{Average order parameters in the mean field model. (a) Average order parameter $V$ in simulations of globally coupled 3d velocity vectors with coupling to noise ratio $\kappa$ subject to correlated noise with coefficient of correlation $q$. Two second order transitions occur, at $q=0, \kappa_{cr}=3$ and at $q=1$, $\kappa_{cr}=-2$ (dashed vertical lines). These transitions are shown in (b) for $q=0$ and in (c) for $q=1$ for different dimensions $d$ (solid lines) and compared to simulations (markers). The cross over of the square root of the velocity correlation $\langle C \rangle = \langle V^2 \rangle_t$ as a function of $q$ from $\langle C\rangle=0$ at $q=0$ to $\langle C\rangle=1$ at $q=1$ is shown in (d) with solid lines and compared to simulations (markers). The theoretical distribution of $C$ over the interval $[-1,1]$ for noise correlation $q=0.5$ and increasing dimensions $d$ is shown in (e). The velocity correlations become narrowly distributed around $q$ for $d\to\infty$, an expression of Moran's theorem in our model.}\label{Fig:Fig02}
\end{figure}
The normalizing factor $Z$ and the average squared order parameter $\langle V^2\rangle_t = \langle C \rangle$ have expressions in terms of the Gauss hypergeometric function $_2F_1(a,b;c;z)$ and the beta function. We write
\begin{eqnarray}\label{Eq:CorrelDist_Z}
    Z(a,b;q) &=& \int_{-1}^{1} (1-C^2)^{b-1}(1-qC)^{-a}\,dC \nonumber \\
      &=& \frac{2^{2b-1}}{(1+q)^{a}}B\left(b,b\right)
      \,_2F_1\left(a,b,2b;\frac{2q}{1+q}\right).~~ 
\end{eqnarray}
Then $\langle C\rangle$ follows from
\begin{equation}\label{Eq:C_avrg}
    \langle 1-qC\rangle = 1-q\langle C\rangle = \frac{Z\left(d,\frac{d-1}{2};q\right)}{Z\left(d-1,\frac{d-1}{2};q\right)}.
\end{equation}
We note that $C$ is distributed on the interval $-1\le C\le 1$ but $\langle C \rangle\ge 0$ for $q\ge 0$. 
The square root of $\langle C \rangle = \langle V^2\rangle_t$ is shown in Fig.\ref{Fig:Fig02}d for $d=2,3$ and $4$ and compared to simulations of the Langevin equations.
At this point we can draw the parallel to the Moran effect. In high dimensions the components of the velocity vectors become independent linear processes. Due to entropic forces, manifest as a linear, noise induced drift to smaller values (Appendix A), the components of $\vec{v}$ are approximately Gaussian normal of variance $1/d$. The stochastic quantity $C=\vec{v}\cdot\vec{v}'$ is equal to the ensemble coefficient of correlation in $d$ realizations of two linear stochastic processes $v_i(t)$ and $v'_i(t)$ with $i=1\ldots d$ subject to correlated white noise. The distribution of $C$ for $d$ pairs of Ornstein-Uhlenbeck processes with correlated noise was found in \cite{Nadrajah2016}. For large $d$ it is similar to \eqref{Eq:CorrelDist}. From the law of large numbers follows Moran's theorem that $p(C)=\delta(C-q)$ for $d\to\infty$, i.e. the ensemble coefficient of correlation for $d\to \infty$ is exactly equal to the correlation of the noise. In Fig.\ref{Fig:Fig02}e we plot the distribution \eqref{Eq:CorrelDist} and observe that it converges to a delta distribution at $C=q$. 
\section{Conclusions}
We have calculated exact distributions for the mean velocity $V=\langle \vec{v}\rangle$ and for the velocity correlation $C=\vec{v}\cdot\vec{v}\,'$, describing the alignment of mean field coupled velocity vectors subject to purely uncorrelated noise $q=0$, purely identical noise $q=1$ and in the absence of coupling $K=0$, respectively. In the first two cases there exist critical coupling to noise ratios $\kappa_{cr}$ where transitions from an isotropic velocity distribution to partial alignment ($q=0$) and from intermittent synchronization to permanent, complete alignment ($q=1$) occur. In the uncoupled case $K=0$ the noise correlation $q$ parameterizes a cross-over between zero order at $q=0$ and complete alignment at $q=1$. We call the resulting directed movement in self propelled particles subject to correlated noise, noise induced swarming. For  $0<q<1$ 
a cross-over from low to high order is observed (Fig.\ref{Fig:Fig02}a,d) but no critical transition when the coupling strength $\kappa$ is changed. In the light of our mean field results it is surprising that increasing the correlation length of the noise in spatially extended systems of self propelled particles, without changing the noise strength, can increase the velocity alignment locally but at the same time inhibit the transition to global directed motion. This effect could have important implications for decentralized control of crowds, or for swarms of artificial agents in random environments.
%
%
%
%
%
%
\bibliography{NOISYNC}
\section*{Appendix A : Stochastic differential equations for the order parameters}
\noindent
Let
\begin{equation}\label{EqSM:Strato_dXi}
    dX_i = \mu_i(\vec{X})dt + \sum_{j} \sigma_{ik}(\vec{X})\circ dW_k
\end{equation}
be Stratonovich stochastic differential equations (SDEs) with uncorrelated Wiener processes, i.e. $dW_idW_j=\delta_{ij}dt$. The equivalent Ito SDEs are obtained by adding the Stratonovich shift
\begin{equation}\label{EqSM:Ito_dXi}
     dX_i = \left(\mu_i + \frac{1}{2}\sum_{jk} \sigma_{jk}\partial_j\sigma_{ik}\right)dt + \sum_k \sigma_{ik}dW_k.
\end{equation}
The Ito SDE for a function $f=f(\vec{X})$ is obtained using Ito's Lemma
\begin{equation}\label{EqSM:Ito_Lemma}
    df = \sum_{i}\partial_i f dX_i + \frac{1}{2}\sum_{ij} \partial^2_{ij}f \, dX_i dX_j
\end{equation}
and together with \eqref{EqSM:Ito_dXi}
\begin{eqnarray}\label{EqSM:Ito_SDE_f}
 df &=& \sum_i \partial_i f \left[\left(\mu_i + \frac{1}{2}\sum_{jk} \sigma_{jk}\partial_j\sigma_{ik}\right)dt + \sum_k \sigma_{ik}dW_k\right] \nonumber  \\
    && + \frac{1}{2}\sum_{ij} \partial^2_{ij}f \sum_{k} \sigma_{ik}\sigma_{jk} \, dt.
\end{eqnarray}
The sum of the increments of the uncorrelated Wiener processes can be cast as the increment of a single Wiener process $\tilde{W}$
\begin{equation}\label{EqSM:Sum_Wi}
    \sum_i \partial_i f \sum_k \sigma_{ik} dW_k = \tilde{\sigma} d\tilde{W}.
\end{equation}
With noise strength
\begin{equation}\label{EqSM:DiffCoeff}
    \tilde{\sigma}^2 = \sum_{ij}\partial_if \partial_j f \sum_{k} \sigma_{ik}\sigma_{jk}
\end{equation}
and the drift term
\begin{equation}\label{EqSM:DriftCoeff}
    \tilde{\mu} = \sum_i \left[\partial_i f \left(\mu_i+\frac{1}{2}\sum_{jk}\sigma_{jk}\partial_j\sigma_{ik}\right)+\sum_{jk}\partial^2_{ij}f\sigma_{ik}\sigma_{jk}\right]
\end{equation}
the Ito SDE for a function $f$ is
\begin{equation}\label{EqSM:Ito_SDE_1d}
    df = \tilde{\mu}dt + \tilde\sigma d\tilde{W}.
\end{equation}
The order of a velocity field is quantified by the alignment of vectors. Under rotational symmetry of a distribution the order parameter in the system becomes a one dimensional stochastic process with $\tilde{\mu}=\tilde{\mu}(f)$ and $\tilde{\sigma}=\tilde{\sigma}(f)$ where the stationary probability density $p(f)$ is current free, i.e.
\begin{equation}\label{Eq:FPE_current_free}
    \tilde{\mu} p = \frac{1}{2}\partial_f \left(\tilde{\sigma}^2 p \right).
\end{equation}
This equation is solved by
\begin{equation}\label{EqSM:FPE_general_sol}
    p(f) = \frac{1}{Z} \frac{1}{\tilde{\sigma}^2} e^{\int^f \frac{2\tilde{\mu}(s)}{\tilde{\sigma}^2(s)}ds}
\end{equation}
where $Z$ is a normalization constant. 
\\ \\
{\bf Case of uncorrelated noise :} Given a force on the velocities $\vec{v}_n$ %
\begin{equation}\label{EqSM:Force_ind_noise}
    \vec{F}_n = \kappa V_0 \vec{e}_z + \sqrt{2}\vec{\xi}_n
\end{equation}
with a fixed deterministic part, here without loss of generality in the $z$ direction, and uncorrelated Gaussian white noise $\vec{\xi}_n$ the Stratonovich SDEs \eqref{EqSM:Strato_dXi} for $\vec{v}$ are
\begin{eqnarray}\label{EqSM:Strato_dvi}
    dv_i &=& \kappa V_0 \left(\delta_{iz} - v_zv_i\right)dt \\
         & & + \sqrt{2} \sum_j \left(\delta_{ij} - v_i v_j \right)\circ dW_j. \nonumber
\end{eqnarray}
It follows
\begin{eqnarray}\label{EqSM:vi_mu_sigma}
    \mu_i &=& \kappa V_0 (\delta_{iz}-v_zv_i) \\
    \sigma_{jk} &=& \sqrt{2}\left(\delta_{jk}-v_jv_k\right) .
\end{eqnarray}
The Ito SDEs \eqref{EqSM:Ito_dXi} for the velocity components are
\begin{eqnarray}\label{EqSM:Ito_dvi}
    dv_i &=& \left(\kappa V_0\left(\delta_{iz} - v_zv_i\right) -(d-1)v_i\right)dt \\
         & & + \sqrt{2}\left(dW_i - v_i\sum_{j}v_jdW_j\right). \nonumber
\end{eqnarray}
Here we note the linear entropic force or noise induced drift of strength $d-1$ towards zero. The Ito SDE \eqref{EqSM:Ito_SDE_1d} for $f=v_z$ follows with Eq.\,\eqref{EqSM:DiffCoeff} and  \eqref{EqSM:DriftCoeff} as
\begin{equation}\label{EqSM:vz_Ito_SDE}
    dv_z = \left[\kappa V_0 (1-v_z^2) - v_z (d-1)\right]dt + \sqrt{2(1-v_z^2)}d\tilde{W}
\end{equation}
and we calculate the stationary distribution \eqref{EqSM:FPE_general_sol}
\begin{equation}\label{EqSM:vz_dist}
    p(v_z) = \frac{1}{Z} (1-v_z^2)^{\frac{d-3}{2}}e^{\kappa V_0 v_z}.
\end{equation}
The normalization constant is a modified Bessel function of the first kind and the first moment is given by the ratio
\begin{equation}\label{Eq:vz_mean}
    V = \langle v_z \rangle = \frac{I_{d/2}(\kappa V_0)}{I_{d/2-1}(\kappa V_0)}.
\end{equation}
In the stationary state $V_0$ must be equal to the average of $v_z$ leading to the self-consistency condition given in the main text. 
\\ \\
{\bf Case of purely identical forcing :}
The force on all velocity vectors is
\begin{equation}\label{EqSM:Force_com_noise}
    \vec{F} = \kappa \langle \vec{v}\rangle + \sqrt{2}\vec{\eta}.
\end{equation}
The parameter $\vec{a}$ of the invariant family of velocity distributions is subject to the Stratonovich SDE
\begin{equation}\label{EqSM:LowDimDyn_a}
    \dot{\vec{a}} = \frac{1}{2}(1+a^2)\vec{F}-\left(\vec{F}\cdot\vec{a}\right)\vec{a}.
\end{equation}
With $\langle \vec{v}\rangle = \langle \Phi_d(a) \vec{a} \rangle$ \cite{chandra2019complexity,Kato2020Cauchy,crnkic2021synchronization} we find the Ito SDE for $a=|\vec{a}|$ observing
\begin{eqnarray}\label{EqSM:a_mu_sigma}
    \mu_i &=& \frac{a_i}{2}(1-a^2)\kappa\Phi_d(a) \\
    \sigma_{ik} &=& \sqrt{2}\left(\frac{1}{2}(1+a^2)\delta_{ik}-a_ia_k\right)
\end{eqnarray}
and after calculating the Ito drift and diffusion terms \eqref{EqSM:DiffCoeff} and \eqref{EqSM:DriftCoeff}
\begin{eqnarray}\label{EqSM:a_Ito_SDE}
    da &=& \frac{1-a^2}{2a}\left(\kappa\Phi_d(a) a^2 + \frac{(d-3)(1+a^2)}{2} +1\right)dt \nonumber \\
    && + \sqrt{2}\frac{1}{2}(1-a^2) d\tilde{W}.
\end{eqnarray}
The stationary distribution \eqref{EqSM:FPE_general_sol} for $f=a$ is
\begin{equation}\label{EqSM:a_dist}
    p(a) = \frac{1}{Z}a^{d-1}(1-a^2)^{-d}\exp\left[\int_0^a \frac{2\kappa\Phi_d(s)s}{(1-s^2)}ds\right].
\end{equation}
The time averaged order parameter $\langle V\rangle_t = \langle \Phi_d(a)a \rangle$ is expressed in terms of
\begin{eqnarray}\label{EqSM:a_dist_Zd}
    Z_d(b,c;\kappa) = \int_0^1 a^{b-1}(1-a^2)^{-c}\exp\left[\int_0^a \frac{2\kappa\Phi_d(s)s}{(1-s^2)}\,ds\right]da. \nonumber \\
\end{eqnarray}
Note that the derivative of the exponential function with respect to $a$ results in a factor $\Phi_d(a)a$ to the probability density. By partial integration we therefore obtain
\begin{equation}\label{EqSM:V_avrg_general}
   \langle V\rangle_t = \frac{(1-d)}{2\kappa}\frac{Z_d(d-1,d-1;\kappa)+2Z_d(d+1,d;\kappa)}{Z_d(d,d;\kappa)}.
\end{equation}
For $d=2$ we have $\Phi_2(s)=1$ \cite{Kato2020Cauchy,crnkic2021synchronization}
and therefore
\begin{equation}\label{EqSM:Zd_2d}
    Z_2(b,c;\kappa) = \int_0^1 a^{b-1}(1-a^2)^{-c-\kappa}\,da = \frac{1}{2}B\left(\frac{b}{2},1-c-\kappa\right).
\end{equation}
For $d=4$ we have $\Phi_4(s)=(3-s^2)/2$ \cite{Kato2020Cauchy,crnkic2021synchronization} and with
\begin{equation}\label{EqSM:Exp_Int_4d}
    \exp\left[\int_0^a \frac{2\kappa\Phi_4(s)s}{(1-s^2)}\,ds\right]
    = (1-a^2)^{-\kappa}e^{\frac{1}{2}\kappa a^2}
\end{equation}
we obtain
\begin{eqnarray}\label{EqSM:Zd_4d}
    &&Z_4(b,c;\kappa) = \int_0^1 a^{b-1}(1-a^2)^{-\kappa-c}e^{\frac{1}{2}\kappa a^2}\,da \\
        &&= \frac{1}{2} M\left(\frac{b}{2},1-c-\kappa+\frac{b}{2};\frac{\kappa}{2}\right) B\left(\frac{b}{2},1-c-\kappa\right) \nonumber
\end{eqnarray}
where $M(a,b;z)$ is Kummer's confluent hypergeometric function.
\\ \\
{\bf Uncoupled case : }
Finally we calculate the velocity correlation $C=\vec{v}\cdot\vec{v}\,'$ for two unit length velocity vectors diffusing on a sphere under correlated noise $\vec{\eta}(t)$ and $\vec{\eta}\,'(t)$ and without mean field coupling, i.e. $\kappa=0$. The velocity correlation $C=C(\vec{v},\vec{v}\,')=\vec{v}\cdot\vec{v}\,'$ is bilinear so that the Ito SDEs for the stochastic process $f=C(t)$ are obtained as
\begin{equation}\label{EqSM:Ito_Lemma_dC}
    dC = \vec{v}\cdot d\vec{v}\,' + d\vec{v}\cdot \vec{v}\,' + d\vec{v}\cdot d\vec{v}\,'
\end{equation}
We use \eqref{EqSM:Ito_dvi} with $\kappa=0$, \eqref{EqSM:Ito_Lemma_dC} together with $dW_idW_j\,' = q\delta_{ij}dt$, calculate the combined diffusion coefficient for a single Wiener process $\tilde{W}$ and find
\begin{eqnarray}\label{EqSM:Ito_SDE_dC}
    dC &=& (-2(d-1)C+2q(d-2+C^2))dt \nonumber\\
       & & + 2\sqrt{(1-C^2)(1-qC)}d\tilde{W}.
\end{eqnarray}
With this one dimensional Ito SDE and Eq. \eqref{EqSM:FPE_general_sol} we find the stationary distribution
\begin{equation}\label{EqSM:CorrelDist}
    p(C) = \frac{1}{Z}\frac{(1-C^2)^{\frac{d-3}{2}}}{(1-qC)^{d-1}}.
\end{equation}
The normalization constant and $\langle C\rangle$ are expressed in terms of the Gauss hypergeometric function $_2F_1(a,b;c;z)$ and the beta function \eqref{Eq:CorrelDist_Z},\eqref{Eq:C_avrg} .
\end{document}